\documentclass[12pt]{article}
\input{epsf.tex}
\def\ba{\begin{eqnarray}}
\def\ea{\end{eqnarray}}
\def\psit{\stackrel{-}{\psi}}
\def\a{\alpha}
\def\b{\beta}
\def\s{\sigma}
\def\r{\rho}
\def\d{\delta}
\def\e{\varepsilon}
\def\g{\gamma}
\def\part{\partial}

\title{\bf
Moving medium electrodynamics approach \\
for description 
of the electric and magnetic static
polarizable properties of the 
nucleon \\ at low energy.}

\author{A.~Ilyichev$^{a}$\thanks{E-mail: ily@hep.by}, 
S. Lukashevich$^{b}$\thanks{E-mail: lukashevich@gsu.unibel.by}, 
N. Maksimenko$^{b}$\thanks{E-mail: maksimenko@gsu.unibel.by}\\
{\small \it $^a $National Scientific	
and Educational Centre of Particle and High Energy Physics}
\\[-1mm]
{\small \it  of the Belarusian State University,
220040  Minsk,  Belarus}\\
{\small \it $^b $Gomel State  University, 246699 Gomel, Belarus}}
\begin{document}
\date{}
\maketitle                % Produces the title.

\abstract{%
Using the  relativistic electrodynamics  of continuous media
formalism  and main relativistic quantum field theory principles
the covariant Lagrangian of electromagnetic field  interaction
with polarizable 1/2-spin  particles have been obtained. This
Lagrangian let us to determine canonical and metric
energy-momentum tensors as well as low-energy Compton scattering
amplitude. The application of this Lagrangian for the calculation of 
the radiative correction to the imaginary part 
of double virtual Compton scattering is demonstrated.}

\section{Introduction} \label{s1}
At present time, the description  of the  Compton scattering off
hadrons is performed by nonrelativistic Hamiltonian function \cite{9,15}.
However for the extraction of the more essential experimental and
theoretical information about 
hadron polarizabilities it is necessary to develop the Lagrangian of
electromagnetic field interaction with polarizable particle
in covariant form.

The such Lagrangian was constructed
by the phenomenological formfactor-based approach
in ref.~\cite{lvov}.
In the present report for the construction of the
similar Lagrangian 
the developed in ref.~\cite{max} approach
is used in more consistently and completely way
that allows us  to  determine not only 
the Lagrangian itself but
the energy-momentum tensor of electromagnetic field  interaction with
polarizable particles  as well as to take into account  introduced
in ref.~\cite{max,rag} spin polarizabilities of hadrons.

One of the most interesting topic in 
Compton scattering investigation is the measurement of
$Q^2$-dependence of 
the forward polarizabilities 
%in deep-inelastic scattering (DIS)
\cite{dd} that can be presented as the imaginary part of
doubly virtual Compton scattering (VVCS) amplitude \cite{sp}.

Performing a such kind of the experiments it will be also important to
take into account radiative effects correctly. Notice that 
FORTRAN codes  have been already developed for estimation of the radiative
effects from lepton legs in elastic (MASCARAD \cite{MASC}) and deep-inelastic
(POLRAD \cite{POLRAD}) scattering. Basing on these codes Monte Carlo generators
for simulation of radiative events in elastic (ELRADGEN \cite{ELRAD}) and deep-inelastic
(RADGEN \cite{RADGEN}) scattering was developed.

Another interesting source of radiative effect contributed to the imaginary part of
VVCS amplitude that considered in the present report
is photon emission from nucleon line. Using the standard Feynman technique
on the lowest order this process can be presented in the frame of the static 
polarizability contributions.

\section{Lagrangian} \label{s2}
The responsible for the interaction of
electromagnetic field with a
structural particle Lagrangian we present in a following way:
\begin{eqnarray}
{\cal L}_I=-{j_\mu}^{(M)}(x)A^\mu(x), \label{a1}
\end{eqnarray}
where ${j_\mu}^{(M)}$ is  the current density due to a motion of
constituents, $A^\mu$ is electromagnetic field four-dimensional
potential.

The current density ${j_\mu}^{(M)}(x)$ have  to satisfy
conservation law
\begin{eqnarray}
\partial^\mu{j_\mu}^{(M)}=0,
\nonumber
\end{eqnarray}
that allows us to to present  it in the form
\begin{eqnarray}
{j_\mu}^{(M)}=-\partial^\rho G_{\rho\mu}^I, \nonumber
\end{eqnarray}
where $G_{\rho\mu}^I$
is an antisymmetric tensor.

 Let us introduce the relativistic generalization of the electric
\begin{eqnarray}
D_{\nu }=G_{\nu \rho }^{I}U^{\rho } \label{a2}
\end{eqnarray}
and magnetic
\begin{eqnarray}
M^{\sigma }=\varepsilon ^{\sigma \rho \kappa \mu }G_{\rho \kappa
}^{I}U_{\mu } \label{a3}
\end{eqnarray}
dipole moment of the system. In the expressions~(\ref{a2})
and~(\ref{a3}) four-dimensional velocity of structural particle
$U$ has components
\begin{eqnarray}U^{\mu }\left\{ \gamma
,{\bf{v}}\gamma \right\} ,
\nonumber
\end{eqnarray}
where $\displaystyle \gamma =\frac{1}{\sqrt{1-{\bf{v}}^{2}}}$, and
${\bf{v}}$ is moving media velocity.

In the rest system of the structural system the vectors $D_\nu$
and $M^\sigma$ can be presented as a  multiple expansion
concerning the center of gravity \cite{5}
\begin{eqnarray}
 D_{k}=\sum_{L=1}^{\infty}(-1)^{L-1}~Q_{kl...n}^{(L)} ~\partial _{l...}~\partial
_{n}~\delta (\overrightarrow{x}),\nonumber
\end{eqnarray}
\begin{eqnarray}M_{k}=\sum_{L=1}^{\infty}(-1)^{L-1}~M_{kl...n}^{(L)} ~\partial
_{l...}~\partial _{n}~\delta (\overrightarrow{x}),\nonumber
\end{eqnarray} where $k,~l=1, 2, 3$; $~Q^{(L)}$ and $M^{(L)}$ are  multiple
tensors, which determined from  a displacement of constituents
with regard to center of gravity.

Directly from (\ref{a2}) and (\ref{a3}) we can find
that $ U^{\nu }D_{\nu
}=0,~~U^{\nu }M_{\nu }=0 $.

Taking into account the space inversion  
one can construct  antisymmetric tensor $G_{\mu \nu }^{I}$
decomposed over $U^{\nu },~~D^{\nu }$ and $M_{\nu }$ vectors 
in a following way:
\begin{eqnarray}
 G_{\mu \nu }^{I}=U_{\mu }D_{\nu }-U_{\nu }D_{\mu }+\varepsilon
_{\mu \nu \rho \sigma }U^{\rho }M^{\sigma }.\label{a4}
\end{eqnarray}
From the other hand the effective Lagrangian ~(\ref{a1}) we can present as
\begin{eqnarray}
{\cal L} _{I}=-j_{\mu }^{(M)}A^{\mu }=(\partial ^{\rho }G_{\rho
\mu }^{I})A^{\mu }. \nonumber
\end{eqnarray}
To avoid ambiguity of the Lagrangian we set
 \begin{eqnarray}
  {\cal L}_{I}=-\frac{1}{2}G_{\rho \mu }^{I}F^{\rho \mu },
\label{a5}
\end{eqnarray}
or by taking (\ref{a4}) into account
 \begin{eqnarray}
{\cal L} _{I}=-\frac{1}{2}(D_{\mu }e^{\mu }-M_{\mu }h^{\mu }),
\label{a6}
\end{eqnarray}
where $e^{\mu },~h^{\mu }$ are express via the electromagnetic
field tensor as $e^{\mu }=F^{\mu \nu }U_{\nu }$, $h^{\mu
}=\widetilde{F}^{\mu \nu }U_{\nu }$.

In the situation when
 \begin{eqnarray}D_{\mu
}=4\pi \alpha _{\mu \nu }e^{\nu }, 
\end{eqnarray}
 \begin{eqnarray}
M_{\mu }=4\pi \beta _{\mu \nu }h^{\nu }, \label{a8}
\end{eqnarray}
and  tensors $\alpha ^{\mu \nu }$ and  $\beta ^{\mu \nu }$  are
represented  in the diagonal  form by using metric tensor $g ^{\mu
\nu }$ as $ \alpha ^{\mu \nu }=g^{\mu \nu }\alpha_0',~~ \beta ^{\mu
\nu }=g^{\mu \nu }\beta_0',$  the Lagrangian ~(\ref{a6})
can be split into two parts:
\begin{eqnarray}
{\cal L}={\cal L}_{0}+{\cal L}_{I},
 \label{a7}
\end{eqnarray}
where $\displaystyle {\cal L }_{0}= -\frac{1}{2}(e^{2}-h^{2})
=-\frac{1}{4}F_{\mu \nu }F^{\mu \nu }=-\frac{1}{4}F^{2}$ and
\begin{eqnarray}
{\cal L}_{I}=-2\pi \left[ (\alpha_0' -\beta_0')~e^{2}+\frac{\beta_0'
}{2}F^{2}\right] .
 \label{a11}
\end{eqnarray}

Inserting the Lagrangian ~(\ref{a7}) into
the Eiler-Lagrange equation of
motion
\begin{eqnarray}
\partial _{\mu }\frac{\partial {\cal L}}
{\partial (\partial _{\mu }A_{\nu })}-\frac{
\partial {\cal L}}{\partial A_{\nu }}=0
 \label{a12}
\end{eqnarray}
one can find
\begin{eqnarray}
\partial _{\mu }F^{\mu \nu }=j^{(M)\nu }=-\partial _{\mu }G^{I \mu \nu },
 \label{a13}
\end{eqnarray}
where
\begin{eqnarray}
 G^{I \mu \nu }=4\pi \left[ (\alpha _0'-\beta _0')(e^{\mu }U^{\nu
}-e^{\nu }U^{\mu })+\beta _0'F^{\mu \nu }\right] .
 \label{a13a}
\end{eqnarray}

From the equations  ~(\ref{a13}) for  media in rest the
definitions of charge and  current densities  for bound charges
are follow
\begin{eqnarray}
\rho ^{(M)}=-4\pi \alpha _0' ({\bf{\nabla }}{\bf{E}})
=-4\pi \alpha _0'\;{\rm div}\;{\bf{E}},
 \label{a14}
\end{eqnarray}
\begin{eqnarray}
{\bf{j}}^{(M)}=4\pi (\alpha _0' \partial _{t}{\bf{E}}-\beta _0'
{\rm rot}\; {\bf{H}}),
 \label{a15}
\end{eqnarray}
while the  Maxwell's equations for media have the form:
\begin{eqnarray}
{\rm rot}\; {\bf{E}}=-\partial
_{t}{\bf{H}},~~{\rm div}\;{\bf{H}} =0,
 \nonumber
 \end{eqnarray}
 \begin{eqnarray}
{\rm rot}\; {\bf{B}}=\partial_{t} {\bf{D}},~~{\rm div}\;
{\bf{D}}=0,
 \label{a16}
\end{eqnarray}
where ${\bf{D}}=\widehat{\varepsilon}{\bf{E}}$ and
${\bf{B}}=\widehat{\mu}{\bf{H}}$  are vectors  of electric and
magnetic induction respectively.

The equations  ~(\ref{a14}) and  ~(\ref{a15})  for media  in rest
can be presented in the covariant form by the introduction  of the
following polarizability tensor of media \cite{3}
\begin{eqnarray}
\widehat{M}=\left(
\begin{array}{cc}
0 & -\bf{P} \\
\bf{P} & \bf{M}^{\times }
\end{array}
\right), \label{a17}
\end{eqnarray}
where ${\bf{P}}$  and ${\bf{M}}$ are electric and magnetic
polarizability  vectors of media. By the definition they are dipole
moments  of volume scale.

 As a result  the charge  and  current densities transform into
\begin{eqnarray}
{\bf{j}}^{(M)}=\partial _t{\bf{P}}-{\rm rot}\;
{\bf{M}},~~\rho ^{(M)}=-{\rm div}\;{\bf{P}}, \label{a18}
\end{eqnarray}
while the equation of motion reads as
\begin{eqnarray}
\partial _\mu (F^{\mu \nu }+G^{I \mu \nu })=j^\nu
\label{a19}
\end{eqnarray}
 for moving media and
\begin{eqnarray}
\partial _\mu (F^{\mu \nu }+M^{\mu \nu })=j^\nu,
\label{a20}
\end{eqnarray}
for media in rest. Here $M^{\mu \nu }$ is the polarizability
tensor of medium (\ref{a17}) and 
 $j^\nu$ is the current density  of  free
charges.

By introduction of the tensor   \cite{5,4,6}
\begin{eqnarray}
G^{\mu \nu }=d^\mu U^\nu -d^\nu U^\mu +\varepsilon ^{\mu \nu \rho
\sigma }U_\rho b_\sigma, \label{a21}
\end{eqnarray}
with $d^\mu =\varepsilon ^{\mu \sigma }e_\sigma ,~~b_\rho =\mu
_{\rho \sigma }h^\sigma, $ the Lagrangian ~(\ref{a2})  for moving
media can be presented as
\begin{eqnarray}
{\cal L}=-\frac 14F^{\mu \nu }G_{\mu \nu }=-\frac 12(e\widehat{\varepsilon }e-h%
\widehat{\mu }h), \label{a22}
\end{eqnarray}
where $\widehat{\varepsilon }=I+4\pi \widehat{\alpha
},~~\widehat{\mu }=I+4\pi \widehat{\beta }.$

\section{The energy-momentum tensor and equation of motion} \label{s3}
The Lagrangian ~(\ref{a2}) helps us to   determine the canonical
energy-momentum tensor. Indeed from the Noether's theorem follows
\begin{eqnarray}
T^{\mu \nu }=\frac{\partial {\cal L}}{\partial (\partial _\mu
A_\rho )}(\partial ^\nu A^\rho )-g^{\mu \nu }{\cal L}. \label{a23}
\end{eqnarray}
Inserting ~(\ref{a22}) into ~(\ref{a23}) one can find that
\begin{eqnarray}
T^{\mu \nu }=-G^{\mu \rho }(\partial ^\nu A_\rho )-g^{\mu \nu
}\frac{1}{4}(F_{\rho \sigma }G^{\rho \sigma }). \label{a24}
\end{eqnarray}

Now we determine the metric energy-momentum tensor as
\begin{eqnarray}
\widetilde{T}^{\mu \nu }=-G^{\mu \rho }(\partial ^\nu A_\rho
)-g^{\mu \nu }L+\partial _\rho (A^\nu G ^{\mu \rho }). \label{a25}
\end{eqnarray}
After  some recombination of ~(\ref{a25})
 using the
equation of motion  ~(\ref{a19}) we find
\begin{eqnarray}
\widetilde{T}^{\mu \nu }=F_\rho \; ^{^{}\nu }G^{\mu \rho }+\frac
14g^{\mu \nu }(F_{\rho \sigma }G^{\rho \sigma }). \label{a26}
\end{eqnarray}
From the relation ~(\ref{a26}) one could find the energy
density for media in rest
\begin{eqnarray}
\widetilde{T}^{00}=\omega =\frac 12(\varepsilon
{\bf{E}}^2+\mu{\bf{H}}^2). \label{a27}
\end{eqnarray}

  Now using the correspondence principle
\cite{1,2} let us consider  the quantum-mechanical description  of
the electromagnetic field interaction with polarizable
particles. The  electric and magnetic polarizabilities extracted
from the Lagrangian  ~(\ref{a22}) can be presented in the form:
\begin{eqnarray}
{\cal L}_I=-2\pi (\alpha _0' F_{\mu \rho }F_{_{}\sigma } \; ^\mu
-\beta _0'\widetilde{F}_{\mu \rho }\widetilde{F}_{_{}\sigma } \; ^\mu
)U^\rho U^\sigma, \label{a28}
\end{eqnarray}
where $\displaystyle \widetilde{F}_{\mu \rho }=\frac 12\varepsilon
_{\mu \rho \sigma \kappa }F^{\sigma \kappa },~~\varepsilon
_{0123}=-1.$

To pass on from  the Lagrangian ~(\ref{a28}) to the Lagrangian of
electromagnetic field  interaction with polarizable
1/2-spin particle we perform a following transition
based on the correspondence principle
between classical and quantum mechanics.
First of all instead
 of $P_\mu$ we introduce a momentum operator acting on the wave
 function of particle $-i{\partial _\mu
}\psi$. Taking into account that $\overline{\psi }\gamma_\mu
{\psi}$ is a current density of particles,   making symmetrization of
operators and providing hermiticity requirement, relativistic and
gauge invariance the Lagrangian of electromagnetic field
interaction with polarizable  1/2-spin particle 
whose mass is equal to $M$ can be
presented  as:
\begin{eqnarray}
{\cal L}_I=\frac{2\pi }M\left[ \alpha_0 F_{\mu \rho }F_{_{}\sigma }
\: ^\mu -\beta _0\widetilde{F}_{\mu \rho }\widetilde{F}_{_{}\sigma }
\; ^\mu \right] \widetilde{ \Theta }^{\rho \sigma }, \label{a31}
\end{eqnarray}
where $\widetilde{ \Theta }^{\rho \sigma }=\displaystyle \frac 12
({ \Theta }^{\rho \sigma }+{ \Theta }^{\sigma \rho })$ is  the energy-momentum tensor
 of spinor field and:
\begin{eqnarray}
\Theta ^{\rho \sigma }=\frac i2\overline{\psi }\gamma ^\rho
\stackrel{\leftrightarrow}{\partial ^\sigma }\psi . \label{a30}
\end{eqnarray}

The polarizabilities $\alpha _0$ and $\beta _0$ introducing in
the expression ~(\ref{a31}) are used in hadronic
physic and connect with the polarizabilities
of expression ~(\ref{a28}) as
\begin{eqnarray}
{\alpha}^{\prime}_0=\rho \alpha_0, ~~ {\beta}^{\prime}_0=\rho \beta_0,
\nonumber
\end{eqnarray}
where $\rho$ is  the  density of
particles.

To verify that the Lagrangian ~(\ref{a31}) is correct it is sufficient
to define Hamiltonian for interaction of electromagnetic field with
polarizable particle and Compton scattering amplitude off this particle.

 Now to  define the moving of the
charged, polarizable, spinor particle  in the electromagnetic
field write out the total Lagrangian in the following form:
\begin{eqnarray}
{\cal L}&=&\frac i2\overline{\psi }\stackrel{\wedge
}{\stackrel{\leftrightarrow}{\partial }} \psi -M\overline{\psi
}\psi -e\overline{\psi }\widehat{A}\psi
-\frac
14F^2-\frac 14F_{\mu \nu }G^{(S) I \mu \nu }. \label{a32}
\end{eqnarray}
Then the tensor  ~(\ref{a13a}) is determined by
\begin{eqnarray}
G^{(S) I \mu \nu }&=&-\frac{4\pi }M\left\{ (\alpha_0 -\beta_0 )\left[
F^{\mu \sigma } \widetilde{\Theta }_{_{}\sigma } \; ^\nu -F^{\nu
\sigma }\widetilde{\Theta } _{_{}\sigma } \; ^\mu \right]
%\right.
%\nonumber\\&&\left. 
+\beta_0 \widetilde{\Theta }_{_{}\rho }
\; ^\rho F^{\mu \nu }\right\} . \label{a33}
\end{eqnarray}
Using the Lagrangian  ~(\ref{a32}) and antisymmetric tensor
~(\ref{a33}) the metric momentum-energy tensor we shall define as
\begin{eqnarray}
\widetilde{T}^{\mu \nu }&=&\widetilde{\Theta }^{\mu \nu }+F_\rho
\; ^{ ^{}\nu }F^{\mu \rho }+\frac 14g^{\mu \nu }F^2
-\frac e2\overline{\psi }(\gamma ^\mu A^\nu +\gamma ^\nu A^\mu
)\psi +\widetilde{T}^{\mu \nu }_{I}, \label{a34}
\end{eqnarray}
where
\begin{eqnarray}
\widetilde{T}^{\mu \nu }_{I}=F_\rho \;  ^{^{}\nu }G_I^{(S)\mu \rho
}+\frac 14g^{\mu \nu }(F_{\rho \sigma }G^{(S) I \rho \sigma }).
\label{a35}
\end{eqnarray}
One can see from the expression  ~(\ref{a35}) that for the
particle in a rest its interaction energy appearing from
polarizability will be have the  form \cite{15}
\begin{eqnarray}
{H}_{I}=-2\pi(\alpha_0 {\bf{E}}^2+\beta_0{\bf{H}}^2). \label{a36}
\end{eqnarray}
 The equation of motion that follows from the Lagrangian~(\ref{a32}) is given by
\begin{eqnarray}
\partial _\mu F^{\mu \nu }=e\overline{\psi }\gamma ^\nu \psi +j^{(M)\nu },
\label{a37}
\end{eqnarray}
where $j^{(M)\nu }=-\partial _\mu G^{(S) I \mu \nu }.$

Taking into account (\ref{a36}) the scattering amplitude  within
second order over photon energy   will give  the contribution  for
electric and magnetic polarizabilities as
\begin{eqnarray} \label{h36}
T_{fi}^{pol}&=&\frac{8\pi M\omega \omega ^{^{\prime
}}}{N(t)}\left[ {\bf e}^{^{\prime }*}\cdot {\bf e} \alpha _0+{\bf
s} ^{^{\prime }*}\cdot {\bf s}\beta _0\right],\label{a38}
\end{eqnarray}
where $ \bf {s}=\bf {n}\times  \bf {e};~\bf {s}^{\prime* }= \bf
{n}^{\prime }\times \bf {e}^{\prime* } ,  $ $\bf {e}$  and $\bf
{e}^{\prime}$ are the polarization vectors, $\omega_{1}$ and
$\omega _2$ are the energies of the incident and scattered
photons, $\bf {n}=\bf {k}/|\bf {k}|$, $\bf {n}^{'}=\bf
{k}^{'}/|\bf {k}^{'}|$.
 
\section{Static polarizability vertex}
For application of the obtained Lagrangian 
to the calculations in quantum field theory 
it is necessary to define the vertex
of photon-nucleon interaction.  
Following the notations of Appendix B of \cite{chl}.
our Lagrangian (\ref{a31})
can be presented as: 
\ba
{\cal L}^{pol}_{eff}&=&
\int \prod_{i=1}^4
\left[ d^4x_i\delta^4(x-x_i) \right ]
\alpha _{\s \d}^{ r'r}(x_1,x_2,x_3,x_4)
\psit_{r'}(x_3)\psi_{r}(x_1)
\nonumber \\&&\times
A^{\s }(x_4)A^{\d }(x_2),
\ea
where $r'r$ are the four-spinor indexes (that usually dropped).
Taking into account that
\ba
\alpha _{\s \d}^{ r'r}(x_1,x_2,x_3,x_4)&=&\int
\frac{d^4p_1}{(2\pi)^4}
\frac{d^4p_2}{(2\pi)^4}
\frac{d^4q_1}{(2\pi)^4}
\frac{d^4q_2}{(2\pi)^4}
{\tilde \alpha }_{\s \d}^{ r'r}(p_1,q_1,p_2,q_2)
\times
\nonumber \\&
\times &
\exp i(p_1(x-x_1)+q_1(x-x_2)-p_2(x-x_3)-q_2(x-x_4)),
\ea
\begin{figure}[t!]
\vspace*{3mm}
\unitlength 1mm
\hspace*{35mm}
\begin{picture}(80,80)
\put(-10,0){
\epsfxsize=12cm
\epsfysize=12cm
\epsfbox{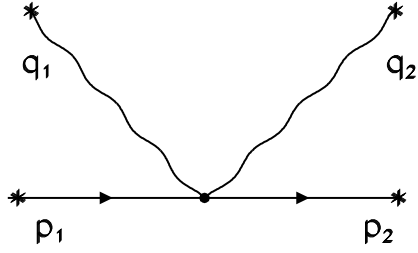}
}
\end{picture}
\vspace*{-50mm}
\caption{
\label{pp}\it
Polarizability vertex
}
\end{figure}

\noindent
where $p_1$ and $q_1$ ($p_2$ and $q_2$)
are the incoming (outgoing) nucleon and photon 
momenta respectively (see Fig.~\ref{pp})
it is easy to show that
\ba
{\tilde \alpha }_{\s \d}^{ r'r}(p_1,q_1,p_2,q_2)
&=&-\frac {\pi}{M}
(p_{1}^{\nu}+p_{2}^{\nu})\g_{\mu}^{ r'r}
(\a_0-\b_0)
[q_{2}^{\mu}\d^{\r}_{ \s}-q_{2}^{\r}\d^{\mu }_{\s}]
[q_{1\r}g_{\nu \d}-q_{1\nu}g_{\r \d}]
\nonumber \\&& 
-\frac 12 \d^{\mu }_{ \nu}\b_0
[q_{2}^{\r}\d^{\g }_{\s}
-q_{2}^{\g}\d^{\r}_{ \s}]
[q_{1\r}g_{\g \d}-q_{1\g}g_{\r \d}]).
\ea
Adding crossing symmetry state
and multiply the result on $i$
we receive the final expression
for polarizability vertex that presented on Fig.~\ref{pp} 
\ba
\Gamma_{\s \d}^{pol}(p_1,q_1,p_2,q_2)&=&i(
{\tilde \alpha }_{\s \d}(p_1,q_1,p_2,q_2)+
{\tilde \alpha }_{\d \s}(p_1,-q_2,p_2,-q_1)
).
\ea
Here we dropped the four-spinor indexes. 
At the end of this section we show the following convolution:
\ba
q_2^{\s}\Gamma_{\s \d}^{pol}(p_1,q_1,p_2,q_2)=0,\;
q_1^{\d}\Gamma_{\s \d}^{pol}(p_1,q_1,p_2,q_2)=0
\ea

\section{Static polarizabilities contributions to VVCS}

\begin{figure}[t!]
%\hspace*{-17mm}
\vspace*{3mm}
\begin{tabular}{cc}
\unitlength 1mm
\hspace*{2cm}
\begin{picture}(80,80)
\put(-30,0){
\epsfxsize=12cm
\epsfysize=12cm
\epsfbox{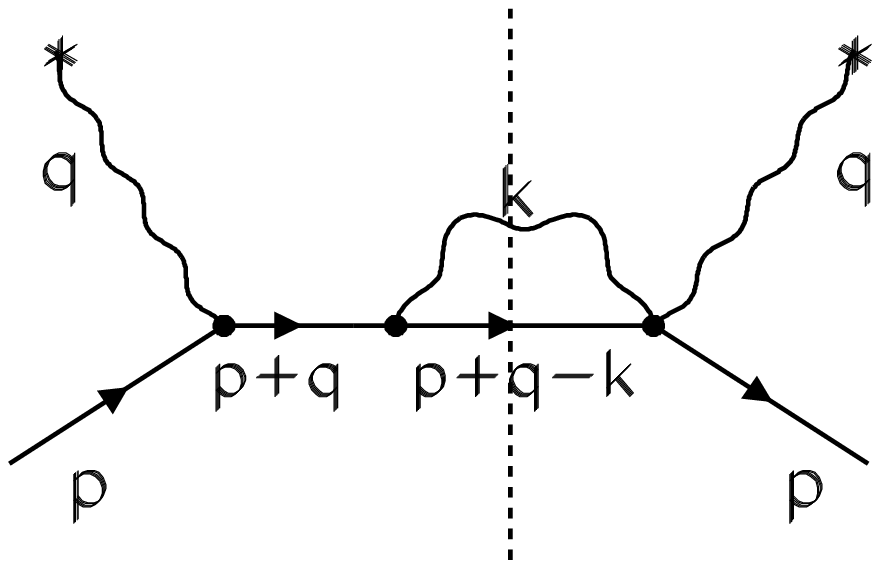}
\put(-65,50){\makebox(0,0){a)}}
}
\end{picture}&
\begin{picture}(80,80)
\put(-130,0){
\epsfxsize=12cm
\epsfysize=12cm
\epsfbox{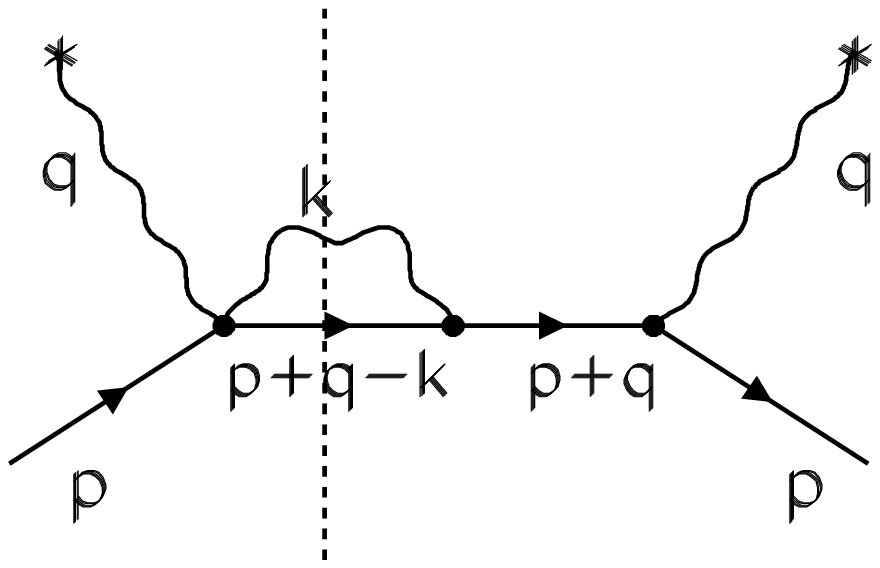}
\put(-180,140){\makebox(0,0){b)}}
}
\end{picture}
\\[15mm]
\hspace*{2cm}
\begin{picture}(80,80)
\put(-160,0){
\epsfxsize=12cm
\epsfysize=12cm
\epsfbox{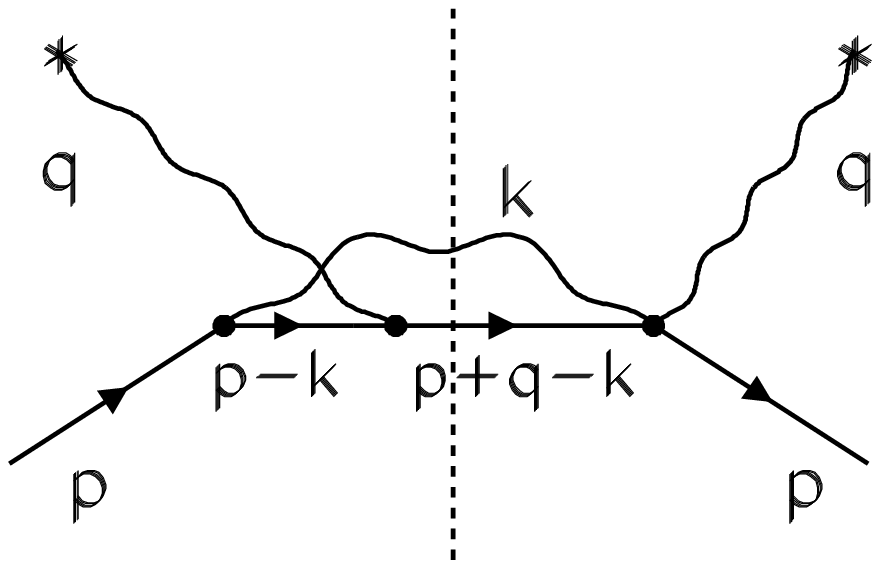}
\put(-175,145){\makebox(0,0){c)}}
}
\end{picture}&
\begin{picture}(80,80)
\put(-130,0){
\epsfxsize=12cm
\epsfysize=12cm
\epsfbox{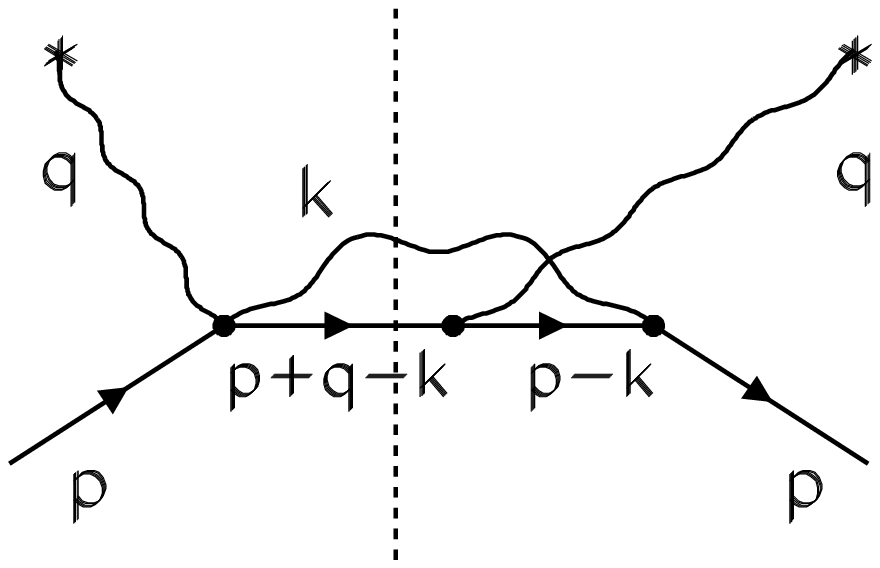}
\put(-180,145){\makebox(0,0){d)}}
}
\end{picture}
\\[13mm]
\hspace*{1cm}
\begin{picture}(80,80)
\put(-50,0){
\epsfxsize=12cm
\epsfysize=12cm
\epsfbox{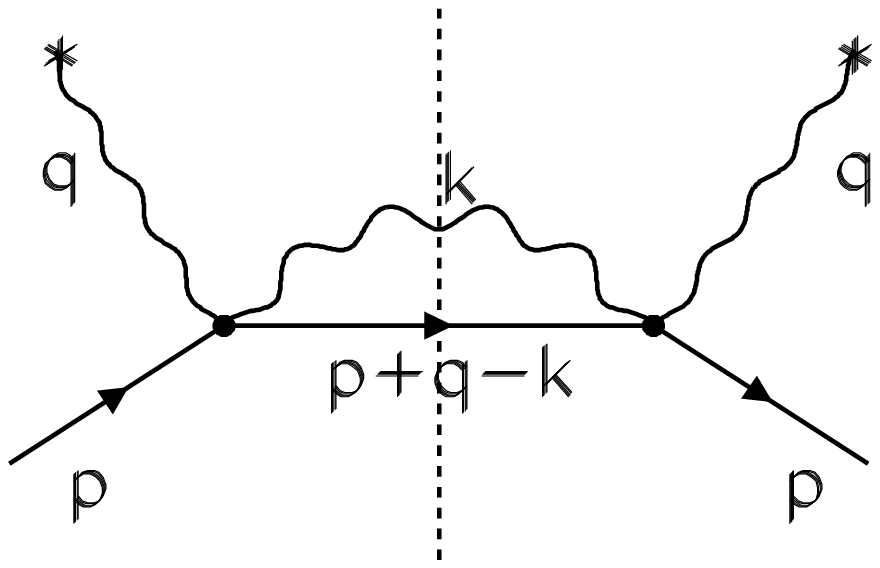}
\put(-180,142){\makebox(0,0){e)}}
}
\end{picture}&

\end{tabular}
\vspace*{-43mm}
\caption{\it
\label{vvc}
Feynman graphs for the lowest order VVCS amplitudes 
whose imaginary parts give contributions 
to the hadronic tensor for inclusive deep-inelastic scattering (DIS). The dashed lines show the cuts 
for imaginary part calculations.   
}
\end{figure}
The lowest order contribution of static polarizabilities to imaginary parts of
VVCS amplitudes and the hadronic tensor of
inclusive DIS as well are presented by Feynman graphs on Fig.~\ref{vvc}. 
 It should be
noticed that contribution from Fig.~\ref{vvc}~(e) is negligible
($\sim 10^{-8}$~fm$^6$) and can be dropped. 
Performing cut over dash line on Fig.~\ref{vvc}~(a-d)
these graph contributions to 
the imaginary part of amplitude can be presented as 
\ba
{\rm Im \;}T
&=&
\pi^2\e^*_{\nu}(q)\e_{\mu}(q)
\int d\Theta
{\bar u}(p)
\left[\frac{
\Gamma^{\mu \nu \;a}_{VVCS}
+\Gamma^{\mu \nu \;b}_{VVCS}
}{(p+q)^2-M^2}+
\frac{
\Gamma^{\mu \nu \;c}_{VVCS}+
\Gamma^{\mu \nu \;d}_{VVCS}
}{(p-k)^2-M^2}\right]
u(p)\nonumber \\
&=&\e^*_{\nu}(q)\e_{\mu}(q)T^{IVVCS}_{\mu \nu}
\label{im}
\ea
where 
\ba
\Gamma^{\mu \nu \; a}_{VVCS}&=&
\Gamma^{pol \;  \nu \a}(p+q-k,k,p,q)
({\hat p}+{\hat q}-{\hat k}+M)
\Gamma_{\a}^{el}(-k)
\nonumber \\[1mm]&&\times
({\hat p}+{\hat q}+M)
\Gamma^{el \; \mu}(q),
\nonumber \\[1mm]
\Gamma^{\mu \nu \; b}_{VVCS}&=&
\Gamma^{el \; \nu}(-q)
({\hat p}+{\hat q}+M)
\Gamma_{\a}^{el}(k)
({\hat p}+{\hat q}-{\hat k}+M)
\nonumber \\[1mm]&&\times
\Gamma^{pol \; \a \mu }(p,q,p+q-k,k),
\nonumber \\[1mm]
\Gamma^{\mu \nu \; c}_{VVCS}&=&
\Gamma^{pol \; \nu \a}(p+q-k,k,p,q)
({\hat p}+{\hat q}-{\hat k}+M)
\Gamma^{el \; \mu }(q)
\nonumber \\[1mm]&&\times
({\hat p}-{\hat k}+M)
\Gamma_{\a}^{el}(-k),
\nonumber \\[1mm]
\Gamma^{\mu \nu \; d}_{VVCS}&=&
\Gamma_{\a}^{el}(k)
({\hat p}-{\hat k}+M)
\Gamma^{el \; \nu}(-q)
({\hat p}+{\hat q}-{\hat k}+M)
\nonumber \\[1mm]&&\times
\Gamma^{pol \;\a \mu }(p,q,p+q-k,k),
\ea
and
$
\Gamma_{\mu}^{el}(q)=-ie
\left (F_D(-q^2)\g_{\mu}+F_P(-q^2)i\s_{\mu \a}q^{\a}/2M\right)
$ 
is the usual elastic vertex.

The phase space  has a form:
\ba 
d\Theta&=&\frac{d^4k}{(2\pi )^4}\delta(k^2)\delta((p+q-k)^2-M^2)
=
\frac{(S_x-Q^2)d\tau d\phi _k}{64\pi^4\tau^2 \sqrt{\lambda _q}}
=\frac{(S_x-Q^2)d\tau}{32\pi^3\tau^2 \sqrt{\lambda _q}},
\ea
where we integrate over an azimuthal photonic angle $\phi_k$.
The invariants have a standard form:
\ba
Q^2=-q^2,\; S_x=2p\cdot q, \; \lambda _q=S_x^2+4 M^2Q^2, \;
\tau=\frac {k\cdot (p+q)}{k\cdot p}.
\ea
The limits of integration over $\tau$ read:
\ba
\tau_{max/min}=1+\frac{S_x\pm \sqrt{\lambda _q}}{2M^2}.
\ea

The contribution of the presented above amplitude to the the
hadronic tensor reads: 
\ba
W^{VVCS}_{\mu \nu}&=&\frac 1{2M\pi}{\rm Tr}
[T^{IVVCS}_{\mu \nu}]
=
\left(g_{\mu \nu}-\frac {q_{\mu}q_{ \nu}}{q^2}\right){\rm Im} \;T_1
\nonumber \\&&
+\frac 1{p\cdot q}\left(p_{\mu }-\frac {p\cdot q}{q^2}q_{\mu}\right)
\left(p_{\nu }-\frac {p\cdot q}{q^2}q_{\nu}\right)
{\rm Im} \;T_2
\nonumber \\&&
+\frac iM \e^{\mu \nu \a \b}q_{\a }\eta_{\b }\;{\rm Im} \;S_1
%\nonumber \\&&
+\frac iM \e^{\mu \nu \a \b}q_{\a }
(p\cdot q \;
\eta_{\b }-\eta \cdot q p_{\b }){\rm Im} \;S_2.
\ea

Using the standard projection operator technique as well as 
the relations
between the imaginary parts of amplitudes and absorption 
cross sections:
\newpage
\ba
{\rm Im} \;T_1=\frac K{4\pi }\sigma _T,\;
%\nonumber \\[2mm]
\qquad
\qquad
\qquad
\qquad
\;\;\;\;
\;\;\;
{\rm Im} \;T_2=\frac {\nu}{M }\frac {Q^2}{\nu^2+Q^2}\frac K{4\pi }(\sigma _T+\sigma _L),
\;\;\;\;
\;\;
\;
\nonumber \\[2mm]
{\rm Im} \;S_1=\frac {\nu M }{\nu^2+Q^2}\frac K{4\pi }
\left (\sigma _{TT}+\frac Q{\nu }\sigma _{LT}\right ),\;
%\nonumber \\[2mm]
{\rm Im} \;S_2=-\frac {M^2 }{\nu^2+Q^2}\frac K{4\pi }
\left (\sigma _{TT}-\frac {\nu }Q\sigma _{LT}\right ),
\ea
where $\nu=S_x/2M $ is a {\it lab } virtual photon energy, 
we can extract $K\sigma_T$, $K\sigma_L$, $K\sigma_{TT}$ and $K\sigma_{LT}$.

\begin{figure}[t!]
\vspace*{-1cm}
\begin{tabular}{cc}
\unitlength 1mm
\hspace*{3cm}
\begin{picture}(80,80)
\put(-30,0){
\epsfxsize=8cm
\epsfysize=8cm
\epsfbox{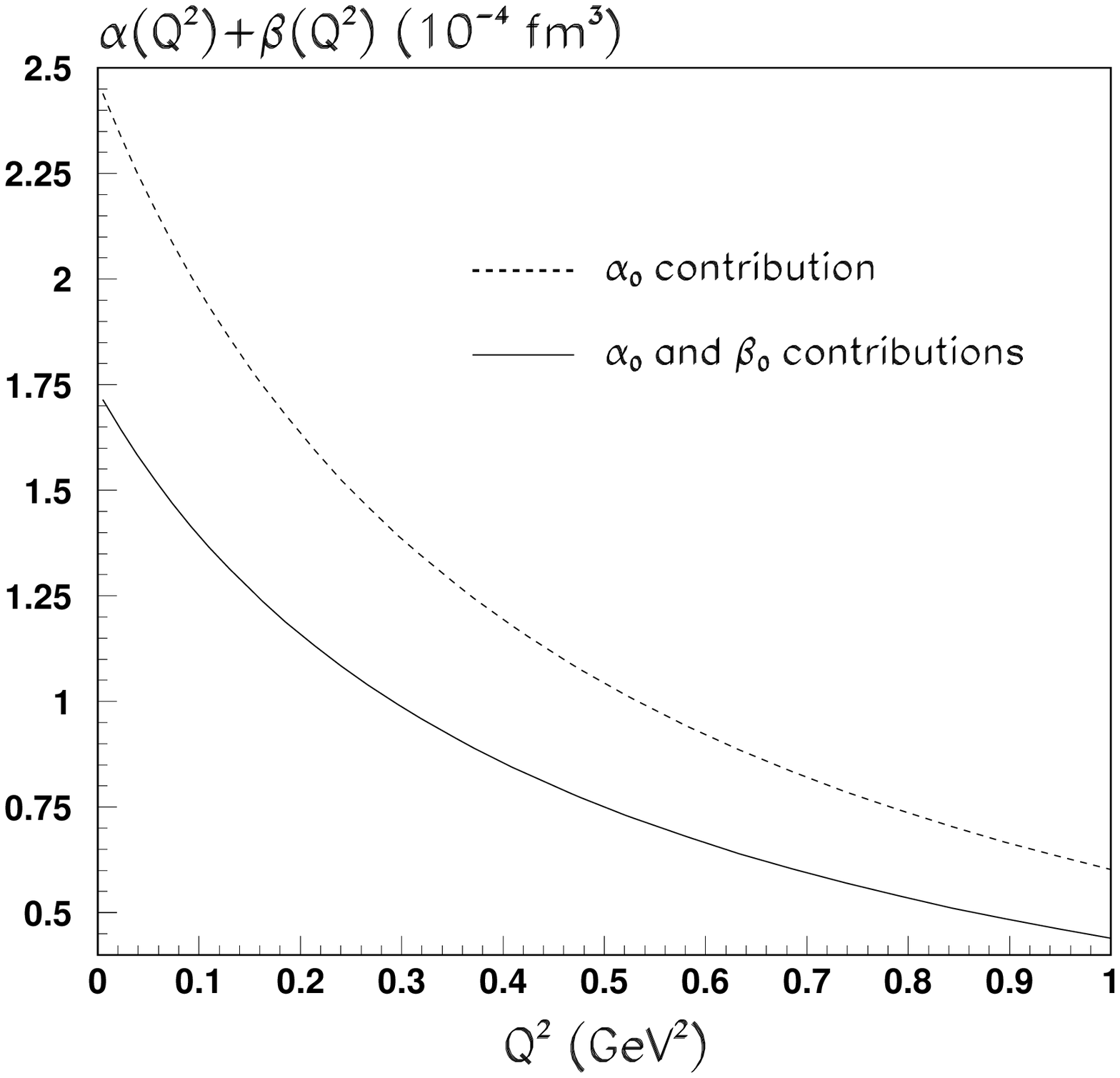}
}
\end{picture}&
\begin{picture}(80,80)
\put(-70,0){
\epsfxsize=8cm
\epsfysize=8cm
\epsfbox{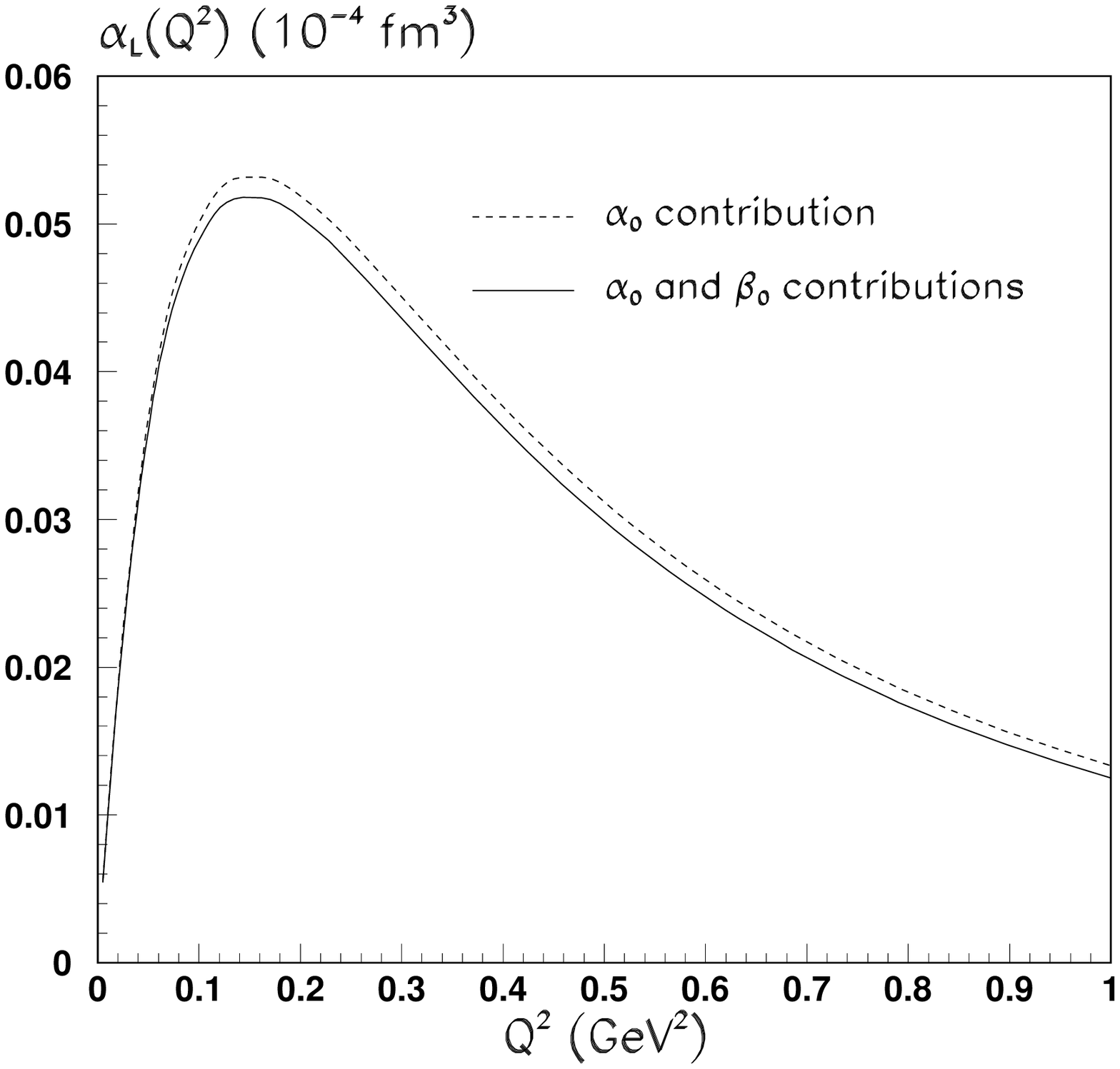}
}
\end{picture}
\\[45mm]
\hspace*{-6cm}
\begin{picture}(80,80)
\put(-30,0){
\epsfxsize=8cm
\epsfysize=8cm
\epsfbox{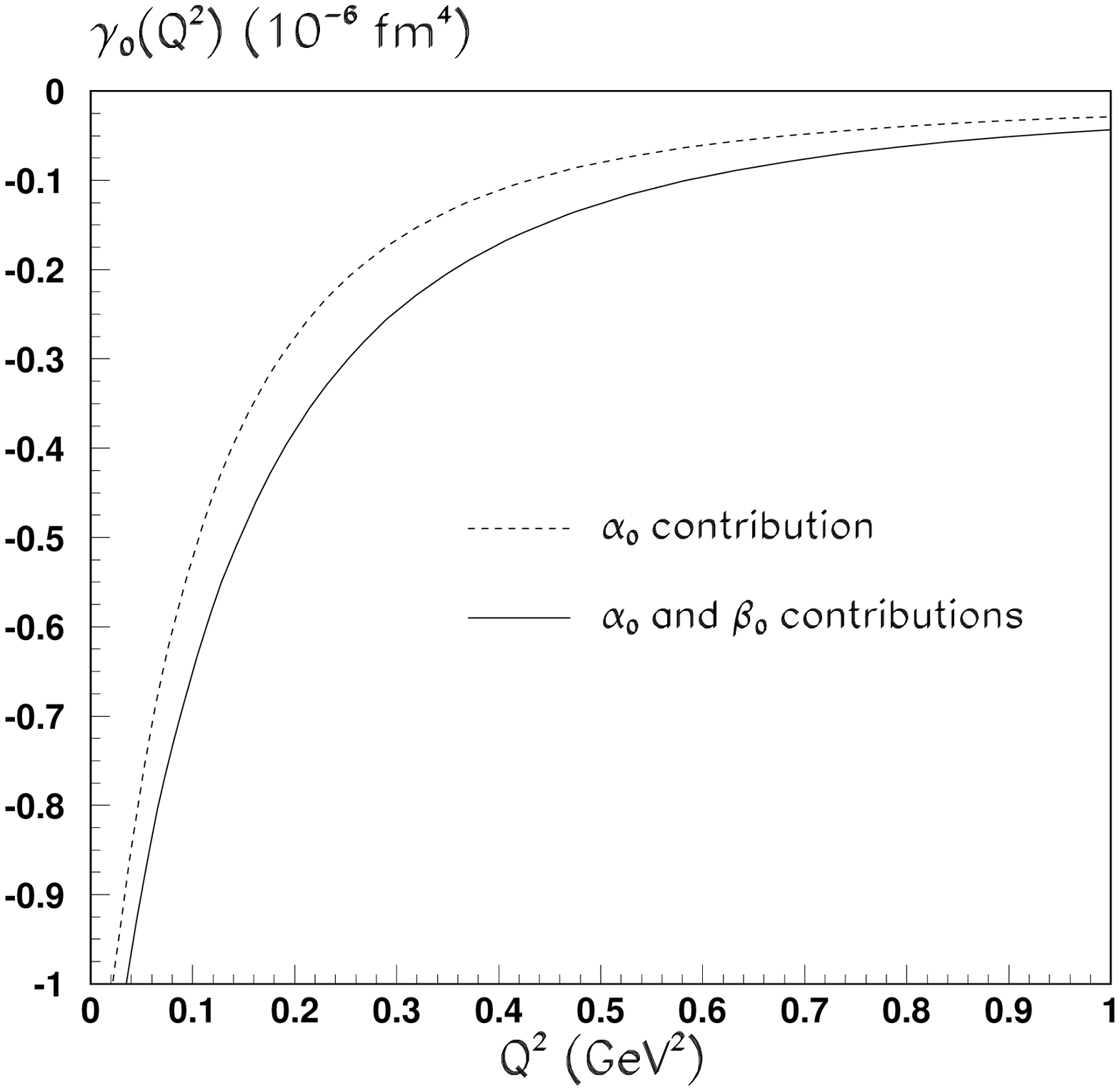}
}
\end{picture}&
\begin{picture}(80,80)
\put(-70,0){
\epsfxsize=8cm
\epsfysize=8cm
\epsfbox{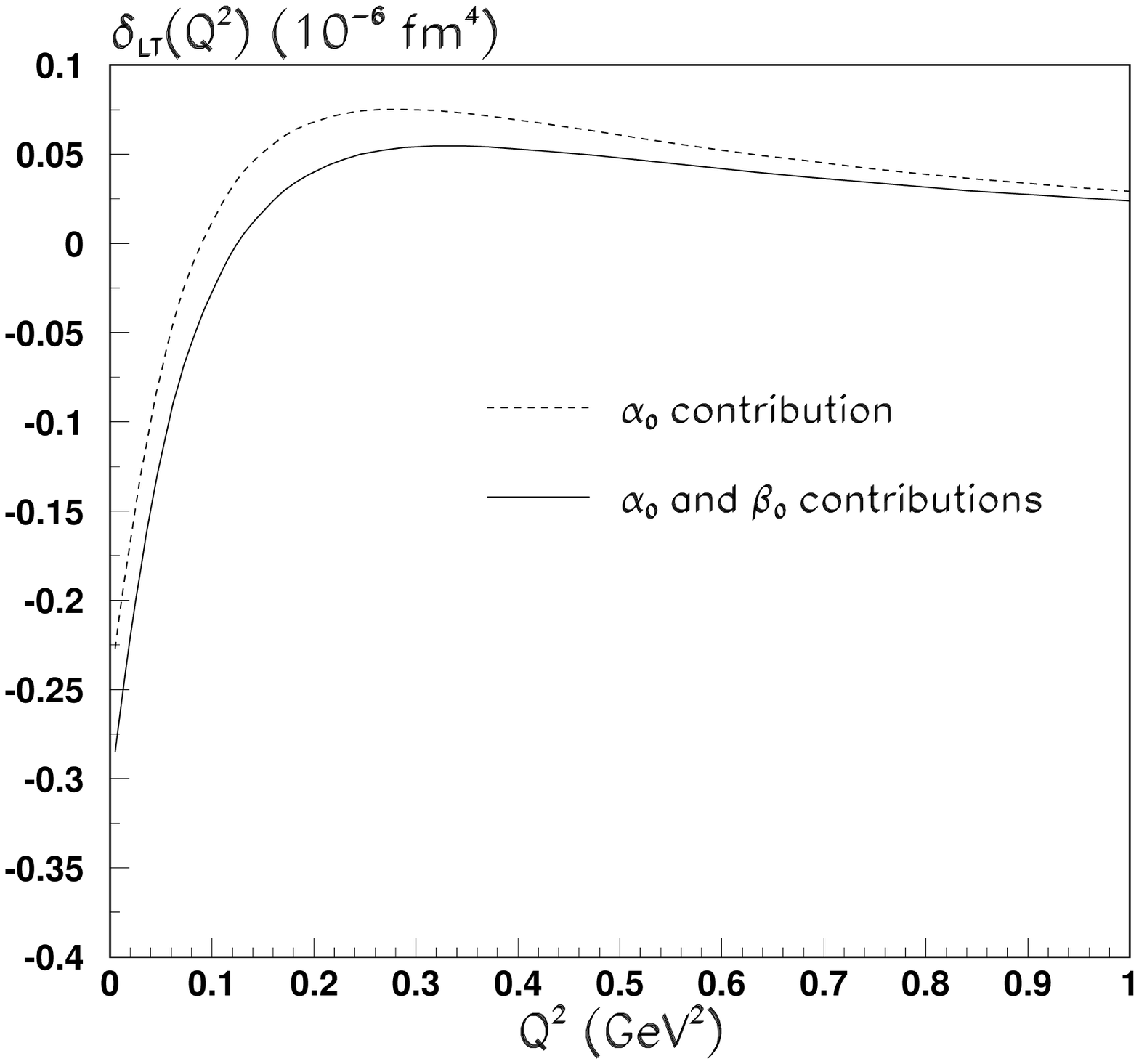}
}
\end{picture}

\end{tabular}
%\vspace*{-43mm}
\caption{\it
\label{gd}
$\a_0$ and $\b_0$ contribution to
$Q^2$ dependence of the generalized forward
polarizabilities of the proton
}
\end{figure}

Taking into account
the asymptotic limit
\ba
\lim _{\nu \gg Q,M} K\sigma_T&=&K\sigma_T^{\nu}=
\frac 8 {3M}\alpha _{QED}\pi^2(\alpha _0-\beta _0)\nu ^3F_P(Q^2)F_P(0),\;
\nonumber \\[2mm]
\lim _{\nu \gg Q,M} K\sigma_L&=&0,
\nonumber \\[2mm]
\lim _{\nu \gg Q,M} K\sigma_{TT}&=&K\sigma_{TT}^{\nu}=
\frac {40} {3M}\alpha _{QED}\pi^2(\alpha _0-\beta _0)\nu ^3F_P(Q^2)F_P(0),\;
\nonumber \\[2mm]
\lim _{\nu \gg Q,M} K\sigma_{LT}&=&K\sigma_{LT}^{\nu}=
\frac {2} {3M}\alpha _{QED}\pi^2\nu ^2Q
((3F_D(Q^2)-2F_P(Q^2))\beta _0
\nonumber \\[2mm]&&%\qquad \qquad \qquad \qquad
-3(F_D(Q^2)-2F_P(Q^2))\alpha _0)F_P(0)
%\nonumber \\[2mm]&&\qquad \qquad \qquad \qquad
-12(\a_0-b_0)F_P(Q^2)F_D(0)),\;
\ea
we can define the contribution of $\a _0 $ and $\b _0$
to the forward polarizabilities in a following way:
\ba
\alpha(Q^2)+\beta(Q^2)&=&
\frac 1{2 \pi ^2}\int \limits_{\nu_0}^{\infty}
K\hat{\sigma}_{T}\frac{d\nu}{\nu^3},
%\nonumber \\[2mm]
\qquad 
\alpha_L(Q^2)=
\frac 1{2 \pi ^2}\int \limits_{\nu_0}^{\infty}
K\sigma_{L}\frac{d\nu}{\nu^3},\;
\nonumber \\[2mm]
\g_0(Q^2)&=&
\frac 1{2 \pi ^2}\int \limits_{\nu_0}^{\infty}
K\hat{\sigma}_{TT}\frac{d\nu}{\nu^4},\;
%\nonumber \\[2mm]
\;\; \d_{LT}(Q^2)=
\frac 1{2 \pi ^2}\int \limits_{\nu_0}^{\infty}
K\hat{\sigma}_{LT}\frac{d\nu}{\nu^3Q},\;
\ea
where 
$K\hat{\sigma}_T=K(\sigma_T-\sigma_T^{\nu})$,
$K\hat{\sigma}_{TT}=K(\sigma_{TT}-\sigma_{TT}^{\nu})$
and $K\hat{\sigma}_{LT}=K(\sigma_{LT}-\sigma_{LT}^{\nu})$.
$Q^2$ dependence of these 
polarizabilities 
for the proton are presented on Fig.~\ref{gd}.

%\vspace*{5mm}
%{\bf Acknowledgments.}  
%The authors would like to thank
%Andrei Afanasev and Jian-Ping Chen for stimulating discussions.

\section{Conclusion}
On the basis of  the  relativistic electrodynamics  of
continuous media formalism and main relativistic quantum field
theory principles the covariant Lagrangian of electromagnetic
field interaction with polarizable 1/2-spin  particles have been
obtained. This Lagrangian let us to determine canonical and metric
energy-momentum tensors as well as low-energy Compton scattering
amplitude.

The present above Lagrangian was apply for calculation of the 
static polarizability contribution to generalized forward
polarizabilities in VVCS. Performed 
numerical analysis shows that:
\begin{itemize}
\item
the electric $\a_0 $ and magnetic $\b_0 $ 
static polarizabilities contribute not only to
$\a(Q^2)+\b(Q^2)$ and $\a_L(Q^2)$
but to $\g_0(Q^2)$ and $\d _{LT}(Q^2)$
as well; 
\item
the behavior of $\a_0$ and $\b_0$ contribution   to
$\a(Q^2)+\b(Q^2)$ and $\a_L(Q^2)$
agrees with the results obtained in other channels and presented by  
\cite{sp};
\item
the behavior of $\a_0$ and $\b_0$ contribution   to
$\g_0(Q^2)$ and $\d _{LT}(Q^2)$
are disagree only at low $Q^2$  with the results obtained 
in other channels and presented by 
\cite{sp};
\item
similar to
\cite{sp}
in real photon limit $Q^2 \to 0$ we found that 
$\a_L(0)=0$ while
$\d _{LT}(0)\ne 0$.
\end{itemize}
In order to estimate the static polarizabilities contribution to 
VVCS correctly 
it is necessary to consider $\gamma _{E1,E2,M1,M2} $ contribution too.
Now basing on
the corresponding principle between the moving medium electrodynamic 
and quantum field theory the Lagrangian with this interaction is under
construction.

\vspace*{3mm}

{\bf Acknowledgments.}  
The authors would like to thank
Andrei Afanasev and Jian-Ping Chen for stimulating discussions.

\end{document}